\documentclass[helvetica, 11pt]{article}
\usepackage{frontiers2008}
\usepackage{times}
\usepackage[english]{babel}
\usepackage[latin1]{inputenc}
\usepackage{graphicx}
\usepackage{hyphenat}
\usepackage{amsmath}
\usepackage{setspace}
\usepackage{multirow}

\begin{document}
\begin{singlespace}
\title{Transactional WaveCache: Towards Speculative and Out-of-Order
  DataFlow Execution of Memory Operations}

\author{Leandro A. J. Marzulo, Felipe M. G. França\\
Universidade Federal do Rio de Janeiro\\ Systems Engineering and Computer Science Program, COPPE \\ Rio de Janeiro - Brazil \\\{lmarzulo, felipe\}@cos.ufrj.br\\
\and
Vítor Santos Costa\\
Universidade do Porto\\
Departamento de Ciência de Computadores\\
Porto - Portugal\\
vsc@dcc.fc.up.pt
}

\maketitle
\thispagestyle{empty}
\end{singlespace}
\begin{abstract}
 The WaveScalar is the first DataFlow Architecture that can
 efficiently provide the sequential memory semantics required by
 imperative languages. This work presents an alternative memory
 ordering mechanism for this architecture, the Transaction
 WaveCache. Our mechanism maintains the execution order of memory
 operations within blocks of code, called Waves, but adds the ability
 to speculatively execute, out-of-order, operations from different
 waves. This ordering mechanism is inspired by progress in supporting
 Transactional Memories. Waves are considered as atomic regions and
 executed as nested transactions. If a wave has finished the
 execution of all its memory operations, as soon as the previous
 waves are committed, it can be committed. If a hazard is detected in
 a speculative Wave, all the following Waves (children) are aborted
 and re-executed. We evaluate the WaveCache on a set artificial
 benchmarks. If the benchmark does not access memory often, we could
 achieve speedups of around 90\%.  Speedups of 33.1\% and 24\% were
 observed on more memory intensive applications, and slowdowns up to
 16\% arise if memory bandwidth is a bottleneck. For an application
 full of WAW, WAR and RAW hazards, a speedup of 139.7\% was verified.
\end{abstract}

\Section{Introduction}

As the speed increase of the single processor slows down, multi-core
machines are becoming standard in environments ranging from the
desktop and portable computing up to servers. This makes parallelism
an important concern, as in order to achieve high performance with the
new architectures one must be able to extract parallelism from the
application. Unfortunately, this is often difficult with the standard
Von Neumman approach to computing, motivating aggressive research on
alternative models such as TRIPS~\cite{Nagarajan2001, 859667} and Raw~\cite{Waingold1997}. A popular
alternative was the \emph{DataFlow model}, where instructions can
execute as soon as they have their input operands
available~\cite{642111, 803050, 2468, 74930,
 325117}, instead of following the program order. Although dataflow
ideas are in use, e.g., in Tomasulo's algorithm~\cite{tomasulo} actual
DataFlow machines were never popular. Arguably, a major problem with a
dataflow was that was hard to support the memory semantics required by
imperative languages. Moving to DataFlow thus required both a new
architecture and a new programming language.

Swanson's WaveScalar architecture is a radical rethought of WaveScalar
concepts that addresses this problem by providing a memory interface
that executes memory accesses according to the program
order~\cite{Swanson2003,SwansonWS, SwansonThesis}. The key idea is
that computation is divided into \emph{waves}, such that each wave
runs as a data-flow computation, but sequencing of waves guarantees
traditional memory access order. It thus becomes possible to run
imperative programs in a dataflow machine and still obtain significant
speedups. Simulation results for WaveScalar show that performance can
be comparable with conventional superscalar designs for single
threaded programs, but with 50\% more area-efficiency. Speedups of 47
on average were achieved for applications with 128
threads~\cite{SwansonThesis}. The tiled organization of the WaveScalar
architecture and the placement algorithms allow an efficient use of
the processor's resources.

In the original Wavescalar design memory requests from different waves
always follow program order. This can be more restrictive than current
designs, where write buffering and read reordering is used
extensively. Work by the WaveScalar group shows that such techniques
can indeed be very useful in the WaveScalar context~\cite{SwansonThesis}. In this work we
present and study an even more aggressive approach. We propose an
optimistic speculative memory access algorithm where we can run
different waves that access memory caches out-of-order. If conflicts
arise, speculative waves are forced to undo their memory accesses. The
approach was inspired by the observation that waves can be regarded as
memory transactions, so the work in Transactional Memories should
apply. Note that although the approach allows more parallelism, it
will requires more complex hardware (on the other hand, such hardware
could also be used to support actual transactions).

In order to obtain detained understanding of our model, we evaluate
performance on a set of artificial applications. Our results show that
performance largely depends on the memory bandwith required by
application, and on the number of hazards found. We first considered
applications with few conflicts. If the benchmark does not access
memory often, we could achieve speedups of around 90\%.  Speedups of
33.1\% and 24\% were observed on more memory intensive applications,
and slowdowns up to 16\% arise if memory bandwith is a bottleneck. For
an application full of WAW, WAR and RAW hazards, a speedup of 139.7\%
was verified.

Section \ref{tm} discusses Transactional Memory concepts that
influenced the creation of Transactional WaveCache. Section \ref{ws}
describes WaveScalar instruction set, memory interface and processor
architecture. Section \ref{TWC} presents the Transactional WaveCache
and Section \ref{res} presents and discusses experiments and
results. Conclusions and future work are presented in Section
\ref{conc}.

\Section{Transactional Memories}\label{tm}

The term \emph{Transactional Memory} (TM) was coined by Herlihy and
Moss \cite{HerlihyM93} as ``a new multiprocessor architecture intended
to make lock-free synchronization as efficient (and easy to use) as
conventional techniques based on mutual exclusion''. The motivation was
to avoid the complexity of lock-based programming. The advent of CMPs
(single-Chip MultiProcessors) has increased demand for easier to
program parallel models and has thus made the subject become popular.

A transaction is an atomic block of code. In TM they can execute
speculatively without blocking. To do so, memory and a register file
must maintain a \emph{record} of all updates by the transaction. In
case there is a conflict between two transactions, one of them is
aborted and re-executed. If a transaction finishes without conflicts,
a \emph{commit} is made. There are four major problems that must be
considered to create a transactional memory solution: data versioning,
conflict detection, nesting and virtualization.

In eager \emph{data versioning} an undo log keeps the backup of memory
addresses and registers changed by transactions and its used to
restore the memory and register file in case of an abortion. If a
commit happens, the undo log is erased. In lazy data versioning a
write buffer keeps the changes made by the transaction. Those changes
are applied to memory and register upon a commit and erased on
abortion. \emph{Conflict detection} can also be eager or lazy. In the
former, conflicts are detected as soon as they happen and in the
later, they are just detected when a transaction finishes (it will
commit if no conflicts are detected).

One further problem in implementing TM is that CPUs may start a new
transactions while still executing a transaction, the so-called
\emph{Nested transactions}. Such transactions may be treated as a
unique transaction (flattening), or may be treated independently
(closed and open nesting~\cite{1136491, nesting-moss-wmpi06}). Last,
transactional memories systems should ensure the correct execution of
programs even when transaction exceeds the scheduler time slice, the
caches and memory capacities or the number of independent nesting
levels allowed by its hardware~\cite{1168903}.

Transactional memories mechanisms may be implemented in hardware
\cite{HerlihyM93, hammond:tcc:ieee-micro:2004, 1116670, 1168903}, where higher performance can be
achieved, but complete solutions to nesting and virtualization
problems may increase the design complexity and make it to
expensive. Software implementations \cite{224987,
 herlihy:stm-dynamic:podc:2003,
 adl-tabatabai:mcrtstm:pldi:2006} allow more complex solutions with
lower performance. Hybrid implementations
\cite{kumar:hybridtm:ppopp:2006, 1168900,
 adl-tabatabai:mcrtstm:pldi:2006} try to get the best of both worlds.

\Section{WaveScalar}\label{ws}

The WaveScalar is the DataFlow architecture used to implement the
Transactional WaveCache, presented in this work. This Section
briefly describes WaveScalar instruction set, memory interface and processor
architecture.

\subsection{Instruction Set}

A DataFlow graph describes a program in the DataFlow model. The nodes
in the graph are \emph{instructions}: instructions are intelligent in
the sense that they have an associated functional unit. Edges
represent the operands exchanged between instructions. The WaveScalar
Instruction Set starts from the Alpha ISA \cite{desikan2001}. The main
difference is that branches must be transformed into a mechanism to
select consumers of values:
\begin{description}
\item[Select ($\phi$):] receives two values $v_1$ and $v_2$ and a
 boolean selector $s$. One of the values $v_1$ or $v_2$ is sent,
 according to $s$.
\item[Steer ($\rho$):] receives a value $v$ and a boolean $b$. The
 value $v$ is sent to one of two paths, depending on $b$.
\end{description}

\subsubsection{Waves}

Waves are connected, acyclic fragments of the control flow graph with
a single entrance. Waves extend hyper-blocks in that they can also
contain joins. The WaveScalar compiler (or binary translator)
partitionates a program into a set of maximal waves. Waves are then
ordered through Wave-ordering annotations.

Notice that different iterations in a loop may execute in parallel in
Dataflow mode, if there are no dependencies between its
instructions. When an instruction in the loop receives an operand
there must be a way to identify to which iteration the operand is
destined. To accomplish that, every operand carries a tag that
indicates the iteration number, or the Wave number. The
\texttt{Wave-Advance} instruction advances the wave numbers for input
operands of a wave. It is associated with every input argument, and is
often merged with the next instruction to reduce overhead.

\begin{figure}[hptb]
\begin{center}
\texttt{
\begin{tabular}{ccc}
\raisebox{3cm}{
	\begin{tabular}{l}
	if(V[0]==0)\\
		\hspace{0.5cm}V[1]=3;\\ 
	else \\ 
		\hspace{0.5cm}V[1]=2;; \\
	V[2]=2; 
	\end{tabular}} & \hspace{1cm} &
\includegraphics[scale=1]{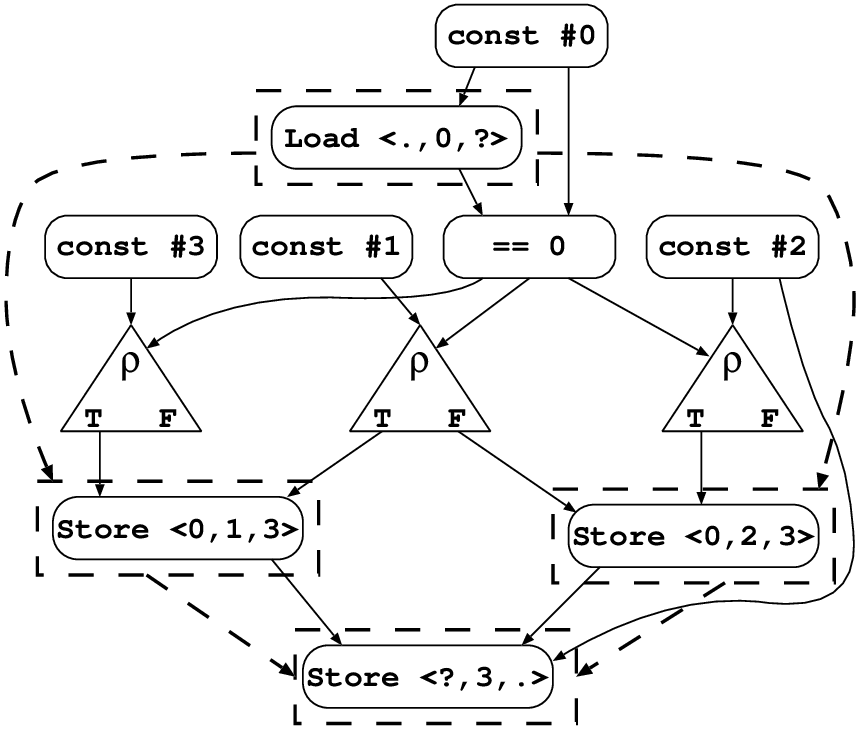} \\
(a) & \hspace{1cm} & (b) \\
\end{tabular}
}
\caption{\label{mem1}DataFlow graph and chain of memory operations in a
 \texttt{IF-THEN-ELSE} block.}
\end{center}
\end{figure}

During compilation, all memory operations further receive a key $<P,
C, S>$ (Predecessor, Current and Successor), that allow the memory
system to establish a chain that connects all memory requests in a
Wave. A memory request can only be executed if the previous request in
the chain and all memory requests from the previous Wave have already
been executed.  When a memory operation is the first or the last of a
Wave, $P$=``.''  and $S$=``.'', respectively (wildcard ``.'' denotes
inexistent operation).  Operations that occur before and after a
branch block have $S$=``?'' and $P$=``?'', respectively (wildcard
``?''  denotes unknown operation). If there are no memory operations
in one of the paths of a branch, there is no way to establish a chain
between the operations that are before and after the branch block. To
solve this problem, a \texttt{MemNop} instruction must be inserted in
that path. Figure \ref{mem1} shows a piece of code with a \texttt{IF-THEN-ELSE} block (a)
and the related DataFlow graph(b). The Wave-ordering annotations 
of Memory operations are also shown, with the dashed lines indicating the
chain that is formed between them.

Wave-ordered memory incorporates simple run-time memory-disambiguation
inside each wave by taking advantage of the fact that the address of a
\emph{Store} is sometimes ready before the data value. When this
happens, the memory system can safely proceed with future memory
operations to different addresses. Stores are broken in two requests:
\emph{Store-Address-Request} and
\emph{Store-Data-Request}. \emph{Store-Address-Requests} arriving
without the corresponding \emph{Store-Data-Request} are inserted in
partial store queues to hold future operations to the same
address. Other operations that access the same address are inserted in
the same queue. Once the \emph{Store-Data-Request} arrives for that
address, the memory can apply all the operations in the partial store
queue in quick succession. Decoupling address and data of Stores
increases memory parallelism by 30\% on average \cite{SwansonThesis}.

\subsubsection{Indirect Jumps}

The \texttt{Indirect-Send} and \texttt{Landing Pad} mechanisms implement
indirect jumps that are used to support function calls with dynamic
linking. The \texttt{Indirect-Send} instruction receives the address
of the target instruction inside the function and a value. The value
is sent to a Landing-Pad that acts as a destination inside the
function. An \texttt{Indirect-Send} is used again to return the
result.

\subsection{The WaveScalar architecture}

\begin{figure}[hptb]
\begin{center}
\includegraphics[scale=0.8]{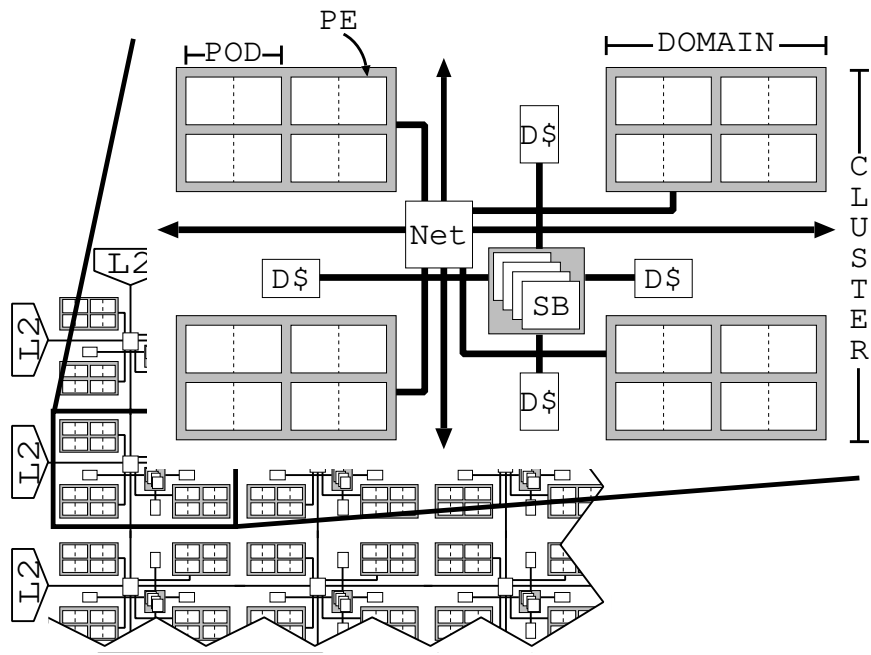}
\caption{\label{waveCache}An Overview of the WaveCache architecture
 (reproduced, with permission of the copyright owner, from~\cite{SwansonThesis})}
\end{center}
\end{figure}

The WaveScalar architecture is called WaveCache and it comprises all
the hardware, except main memory, required to run a WaveScalar
program. It is designed as a scalable grid of identical dataflow
processing elements, the \emph{wave-ordered memory hardware}, and a
hierarchical interconnect to support communication. The
\emph{Cluster}, each one with four \emph{Domains}, is the construction
block of the WaveCache. It has a L1 cache, the \emph{StoreBuffer} that
interfaces with the Wave-ordered memory, and a \emph{Switch} to
provide intra and inter-Cluster communication. Each \emph{Domain} has
eight processing elements grouped in \emph{Pods} of two PEs each. The
\emph{Clusters} are replicated across the die, forming a matrix that
is connected to the L2 cache. Figure~\ref{waveCache} shows an overview
of the WaveCache.

The PEs implement the DataFlow firing rule and thus execute
instructions. Each PE has an ALU, memory structures to hold operands,
logic to control execution and communication, and an instruction
buffer. When a program executes in WaveScalar, multiple instructions
are mapped to the same PE by a placement algorithm. As the program
evolves, some instructions become unnecessary and are replaced by new
ones. The firing rule ensures that an instruction is executed only
when all its operands are available. Every PE has a matching table\label{MatchingTable}
that holds all operands destined to instructions mapped to that PE,
until the instructions is ready to be fired, causing those operands to
be consumed and possibly producing result operands.

\label{WaveMap}
Each Cluster has a \emph{StoreBuffer} (\emph{SB}) that is responsible
for the memory ordering mechanism of some Waves. The Wave Map holds
the information about which \emph{StoreBuffer} has the custody of
which Wave. This Map is stored in main memory; a unique view of the
lines used by each \emph{SB} is guaranteed by the cache coherence
protocol. Memory Requests are sent from the PEs to their local
\emph{StoreBuffers} and routed, if needed, to the remote \emph{SB}
that is responsible for the PE's Wave. The \emph{SB} then inserts the
request in a list for that wave. That request will be executed as soon
as all the previous waves have executed all their memory operations,
and also when a chain is established between the previous executed
memory operation and the present one.

\Section{The Transactional WaveCache}\label{TWC}

By default, WaveScalar follows a strict ordering mechanism for memory
accesses. Partial Stores can add parallelism to the execution of
memory operations within a wave, but parallel execution between waves
if only possible if latter waves do not need memory. The motivation
for our work is that more parallelism becomes available if we allow
disjoint memory operations to execute in parallel.

This work presents an alternative memory ordering mechanism that
maintains the execution order of memory operations within a wave, but
adds the ability to speculatively execute, out-of-order, operations
form different waves. This ordering mechanism is inspired on the way
Transactional Memories works. Waves are considered as atomic regions
and executed as \emph{nested transactions}. In a nutshell, if a wave
has finished the execution of all its memory operations, it can commit
as soon as the previous waves have committed. If an hazard is detected
in a speculative wave, all the following Waves (children) are aborted
and re-executed.

\subsection{Transactions in WaveScalar}

The large body of work in TM suggests a number of alternatives to
implement our approach. Next, we present our algorithm.  The first
idea is that each Wave has a \emph{F} bit that indicates whether it
has finished all its memory operations. This bit is stored in the
\emph{Wave Map} (see page \pageref{WaveMap}). Moreover, each
\emph{StoreBuffer} has an extra attribute called
\emph{lastCommittedWave} that holds the number of the last committed
wave. The \emph{F} bit and the \emph{lastCommittedWave} indicate the
transactions state:

\begin{enumerate}
\item Wave $X$ is non-speculative if all previous Waves are
 committed, i.e., \[ X = lastCommittedWave+1\]
 \noindent The transactional context does not need to be kept for
 non-speculative Waves;
\item Memory operations belonging to a speculative Wave are called
 speculative operations;
\item At initialization
 $StoreBuffer.lastCommittedWave = -1$
 \noindent as the first Wave ($Wave.Id = 0$) is always
 non-speculative;
\item At initialization, $Wave.F = FALSE$;
\item When a wave finishes the execution of all its memory
 operations, $Wave.F=TRUE$;
\item A Wave $Wave$ is pending if
 \[ Wave.Id > lastCommittedWave+1 \land Wave.F\]
\item A Wave $Wave$ can commit when
 \[ Wave.Id = lastCommittedWave+1 \land Wave.F\]
\end{enumerate}

In contrast to standard TM, transaction size is not a problem here:
transactions that need more resources than available should just stall
until resources become available or until they become non-speculative.

\subsection{The transactional context}

In Transaction Memory for Von Neumman machines, before a transaction
starts it is necessary to save the content of all registers used by
the program. A history of all changes in memory must also be kept. If
a hazard is detected this information will be used to restore the
memory and register file. In WaveScalar there is no register file but
all operands used by a transaction that were produced outside the
transaction need to be re-sent. Those operands form the \emph{read
 set} of the transaction or Wave.

We observe that the \texttt{Wave-Advance} instructions control how
operands reach a wave. In order to keep the read set for each Wave, a
natural step is to modify those instructions so that they will send a
copy of their operand to the \emph{StoreBuffer}, that will in turn
insert them in a novel structure called \emph{Wave-Context-Table
 (WCT)}. A StoreBuffer may have several WCTs: each WCT holds the read
set for a wave, such that each line hold a \texttt{Wave-Advance}
operand. If the WCT fills up, the Wave-Advance operands remains in the
PEs output buffers and block wave advance.

We further need to keep a memory log: this is implemented through a
table called \emph{MemOp-History (MOH)}. Per each operation, the MOH
stores the Wave number, the $Current$ number of the operation's memory
annotation, the operation type, the data backup for \texttt{Store}
operations and whether it is a \texttt{Load} or \texttt{Store}. Each
\emph{StoreBuffer} has an associated MOH. Note that one problem is
that waves belong to different SBs can interefere, all the involved
SBs would need a unique view of the MOH so that hazards between waves
could be detected: the MOH would need to reside in memory bringing
high communication costs. We avoid this by ensuring that a speculative
wave must use the same SB as its closest non-speculative
wave. Arguably, this solution could limit parallelism, but in fact
waves in the same thread are usually under the responsibility of the
same SB~\cite{SwansonThesis}.

\subsection{Speculation and Re-execution}

We used the two previous data-structures so that the WaveScalar's
original memory ordering mechanism would allow memory requests from
different waves to be execute concurrently. In order to study how
speculation can affect performance, we can impose a limit on the
number of speculative waves that may execute. We call this limit the
\emph{Speculation Window}.

Before we go on, we observe that problems arise because an instruction
may receive operands from old and recent executions. In the worst
case, operands might match in the \emph{Matching Table (MT)} (see
\pageref{MatchingTable}) and be consumed producing wrong results. We
further add an execution number $ExeN$ to the operands' tags and
change the firing rule so that $ExeN$ is also used to match
operands. An instruction that executes with operands of $ExeN=K$
produces operands also within $ExeN=K$.

The \emph{StoreBuffer} maintains a $lastExeN$ attribute that holds
the number of the last execution.  Waves have also have an $CurrentExeN$
 attribute: a memory request can only be accepted by Wave $Wave$ if 
$ExeN \leq Wave.CurrentExeN$.

We can now describe the rollback algorithm. If a hazard is found in
Wave $Wave$, the WCTs for all waves $Y$ such that $Y.Id> Wave.Id$ are
discarded, causing the re-execution of all following waves; all
operands in the WCT for $Wave$ are then re-sent. Note that at the
moment of a re-execution, some operands that belong to the read set of
\emph{X} might not be present in the WCT yet. This is not a problem:
they simply did not reach the \texttt{Wave-Advance} instruction. When
they do reach: \textbf{(i)} the operands will be inserted in the WCT,
\textbf{(ii)} will have their \emph{ExeN} updated, and \textbf{(iii)}
they will be sent to the new instance of wave $Wave$.

Note that when operands are re-sent by the \emph{StoreBuffers} they
have their \emph{ExeN} changed to a unique number that identifies the
execution. More precisely, when Wave \emph{Wave} causes a
re-execution, all waves $\geq Wave$ have their \emph{currentExeN} set
to $\emph{StoreBuffer}.lastExeN$. Thus, memory requests from old
executions will not be accepted by the memory system.

\subsection{Hazard Detection}

A hazard may exist between two memory operations if they access the
same memory address. Note that as the Transactional WaveCache
maintains Wavescalar memory ordering within waves, operations in the
same Wave will never cause a hazard. When an memory operation belonging 
to Wave $Wave$ is executed, an hazard can only occur between
Wave $Wave$ and some other wave $Y$ such that $Wave.Id > Y.id$.

In our architecture, the MOH structure is the key for recognizing
hazards. More precisely, considering two memory operations $A$ and $B$,
and their respective waves $X$ and $Y$, such that $X.Id > Y.Id$. We
assume that there were no previous speculative stores between them and
that $A$ has executed. When $B$ finally executes, the possible hazards
and respective solutions are as follows:
\begin{description}
\item {\textbf{RAW (Read After Write)}}: Occurs when $A$ is a
  \texttt{Load} and $B$ is a \texttt{Store}. All waves $\geq X$ must
  be aborted and re-executed. If $B$ is speculative, it is inserted in
  the MOH and its backup field is either \textbf{(i)} copied from
  $A$'s value field in the MOH or \textbf{(ii)} obtained from a
  \texttt{Load}.
\item {\textbf{WAW (Write After Write)}}: When both $A$ and $B$ are
  \texttt{Stores}, the Wave $X$ doesn't need to be re-executed. If $B$
  is speculative, it is inserted in the MOH and its backup field
  receives $A$'s backup. Either way, the \texttt{Store} needs not to
  go to main memory, and $A$'s backup field in the MOH receives the
  value of \texttt{Store} $B$.
\item {\textbf{WAR (Write After Read)}}: When $A$ is a \texttt{Store}
  and $B$ is a \texttt{Load}, we need not to re-execute $X$. Instead,
  the backup field of the \emph{A} can be the return value for the
  \texttt{Load}. If $B$ is speculative, it should be inserted in the
  MOH.
\end{description}

\subsection{Committing or re-executing a Transaction}

Commits execute whenever a non-speculative wave finishes its execution
in the memory system. A commit in Wave $X$ makes wave $X+1$
non-speculative. Therefore, the transactional context of $X+1$ can be
erased; if $X+1$ completed (i.e., $F = \mathtt{TRUE}$), it will commit
and we can proceed to the next wave. Note that a Wave can commit and
the next Wave be under the responsibility of a different
\emph{StoreBuffer}. If so, a message is sent to that \emph{SB} so that
it knows that now it has the non-speculative wave and that it can start
executing memory operations. The $lastExeN$ attribute is also sent and
updated in the destination \emph{SB}, to be used in case of future
re-executions.

Committing requires cleaning the MOH. One possibility would be to scan
the whole structure. Our prototype uses a \emph{Search Catalog} to
speedup this operation. For each wave, the Search Catalogue points to
the list including all operations for the wave.

We are now in a position to present the necessary steps to re-execute
a program if wave $X$ was undone:
\begin{enumerate}
\item Erase the lines where $Wave > X$ in the Search Catalog;
\item Restore the memory to its state (using the backup field of
  \texttt{Stores} MOH) before the execution of all operations of waves
  $\geq X$. During this process the MOH for those waves must also be
  erased;
\item Clean the WCTs for all $ Waves > X$;
\item Clean all requests for $Waves \geq X$ in the local
  StoreBuffer. Request in remote SBs will be cleaned lazily when they
  receive requests for the new executions;
\item Increment the $lastExeN$ attribute in the local
  \emph{StoreBuffer} and copy that value to the $currentExeN$ of all
  $Waves \geq X$;
\item Re-send all the operands in $X$'s read set.
\end{enumerate}

\subsection{Erasing Old Operands}

One problem with speculative execution is that the operands from
old executions compete for resources with operands from the current execution. In the worst case,
old operands may find their way to memory access instructions, and
generate memory requests that will not be responded by the
memory system. To make things worst, those operands will not be consumed
from the PE queues and may eventually spill into memory, further
degrading system performance. We propose two mechanisms to address
this problem by removing older operands from processing elements:
the \emph{MT Checkups} and the \emph{Execution Maps}.

\emph{MT Checkups} scan received operands in the matching tables in order to find all
operands to the same instruction, wave, thread and application, but
with different $ExeN$. Operands from older executions should be
erased, whereas the newer operand is inserted in the matching
table. If the received operand is older, it is ignored. If their
execution is the same, the operand is just inserted in the matching table.

For operands that are the first to arrive in a instruction, since
there are no other operands to compare to, \emph{MT Checkups} will not
be useful. Also, old executions are usually ahead of new ones, and the
former will just be reached if it gets stuck in memory accesses
instructions. \emph{Execution Maps} give the PEs the ability to
eliminate operands based on its local view of the execution. A table
containing pairs of $<Wave, ExeN>$ is kept in each PE to hold
information on allowed operands. If a PE has the pairs $<0, 0>$ and
$<5, 1>$ in its Execution Map, that means that from Wave $0$ to Wave
$4$ operands with $ExeN\geq 0$ are accepted and, for waves $\geq 5$
only operands with $ExeN \geq 1$ are accepted. If an operand with Wave
number $3$ and $ExeN=1$ arrives, it will be accepted and the pair $<
5, 1>$ will be replaced by the pair $<3, 1>$ in the Execution Map.

It is important to observe that those mechanisms do not guarantee that
all old operands are going to be erased from the PEs, but they can
minimize the explosion of parallelism caused by the Transactional
WaveCache.

\Section{Experiments and Results}\label{res}

\subsection{Methodology}

Currently, WaveScalar programs must be compiled using the Alpha Tru64
\texttt{cc} compiler and then translated to WaveScalar assembly, using
a binary translator. The binary translator does not support full Alpha
assembly yet, hence most programs are partially run as WaveScalar, and
partially run as Alpha programs. In this work, we extended the Kahuna
WaveScalar architectural simulator~\cite{simuladorws} to support the
Transactional WaveCache. Our changes do not yet support Alpha
emulation, multi-threaded execution and Decoupled Stores.

In order to obtain detained understand of the system, our initial
experiments use a set of kernel benchmarks. The \texttt{MATRIX} group
of benchmarks calculates the determinant of a list of 500 matrices and
also summarizes each line those matrices, filling 500 vectors. In the
\texttt{MATRIX} and \texttt{MATRIX-DEP} applications the matrices are
not initialized. In \texttt{MATRIX-STORES},
\texttt{MATRIX-STORES-DEP}, \texttt{MATRIX-STORES-MIN} and
\texttt{MATRIX-STORES-MIN-DEP} the matrices are initialized before the
loop that calculates the determinants and lines summation, but in
\texttt{MATRIX-STORES-MIN} and \texttt{MATRIX-STORES-MIN-DEP} programs
those calculations are repeated 10 times. The \texttt{MATRIX-DEP},
\texttt{MATRIX-STORES-DEP} and \texttt{MATRIX-STORES-MIN-DEP}
applications have a dependency that may cause RAW hazards between
iterations 249 and 250 of the calculation loop. The
\texttt{VECTOR-FULL-DEP} application executes a loop full of Loads and
Stores in a vector of 500 elements. This is done in a way that forces
WAR, WAW and RAW hazards to be detected.

Table \ref{parq} shows the architectural parameters used in the
simulations. All the applications were executed in the original
WaveScalar, without any memory disambiguation mechanisms and also with
Decoupled Stores. They were also executed using Transactional
WaveCache, with Speculation Window sizes of 2, 3, 5, 10, 20, 30 and
with an infinite Speculation Window. To verify the correctness of the
Transactional WaveCache the final memory results for every simulation
were compared with the original WaveScalar's results. The
Transactional WaveCache structures (WaveContext-Tables, Search
Catalogs, MemOp-Histories and Execution Maps) where not limited in the
implementati

\begin{table}[htbp]
\footnotesize
\centering
\begin{tabular}{|l|p{8cm}|} \hline
Number of Clusters & 4 (2x2) \\ \hline
Domains per Cluster & 4 (2x2) \\ \hline
Processing Elements per Domain & 8 (2x4) \\ \hline
Placement algorithm & Exp2 (dynamic) \\ \hline
\multirow{3}{*}{Cache L1} & Direct Mapping \\
& 1024 lines of 32 bits \\
& Access time: 1 cycle \\ \hline
\multirow{3}{*}{Cache L2} & 4-way set associative \\
& 131072 lines of 128 bits \\
& Access time: 7 cycles \\ \hline
Memory access time & 100 cycles\\ \hline
\multirow{5}{*}{Processing Elements} & Operand queues size: 10000000 lines\\
& Queues search time: 0 \\
& Executions per cycle: 1 \\
& Input operands per instruction per cycle: 3 \\
& Number of instructions: 8 \\ \hline
\multirow{2}{*}{StoreBuffers} & 4 input ports \\
& 4 output ports \\ \hline
\end{tabular}
\normalsize
\caption{\label{parq} Architectural parameters used in the simulations }
\end{table}

\subsection{Results}

\begin{figure}[p]
\begin{center}
\includegraphics[scale=0.8]{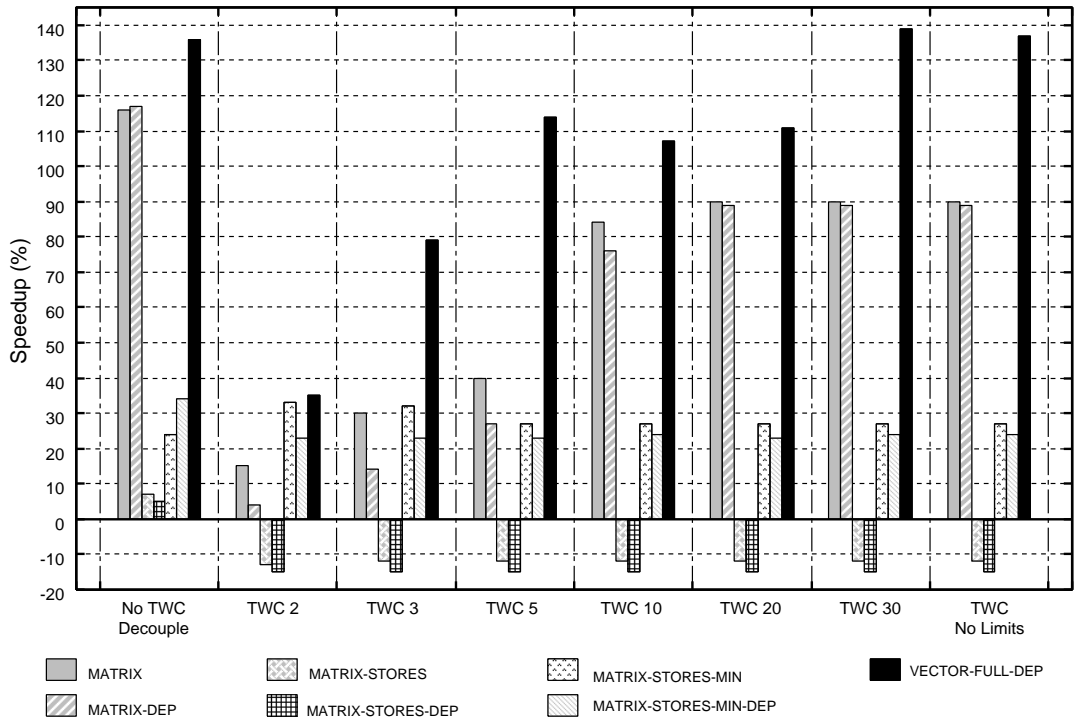}
\caption{\label{toys_speedups} Speedups based on the original WaveScalar}
\end{center}
\end{figure}

Figure \ref{toys_speedups} shows the speedups for all applications
compared to the original WaveScalar without memory
disambiguation. Table \ref{haz} show the number of RAW, WAR and WAW
hazards detected during the simulations. Most of the applications show
a significant speedup that grows with the Speculation Window, although
the maximum speedup is reached for finite Speculation Windows that are
usually small (less then 20).

Decoupled Stores achieves better performance for all applications
except \texttt{VECTOR-FULL-DEP}. For applications with high speedups in the
Transactional WaveCache, there were usually high speedups in Decoupled
Stores. Applications with lower speedups or slowdowns in the
Transactional WaveCache, had also low speedups in Decoupled
Stores. Since both techniques are orthogonal, a combination of two is
possible and is subject of ongoing work.

\texttt{MATRIX-STORES} and \texttt{MATRIX-STORES-DEP} presented
slowdowns of 12.7\% and 15.8\%. This happened because in the
initialization loop there are lots of Stores and few arithmetic or
Loads operations. Stores need two operands to be fired (data and
address), and the addresses are based on the loop control variable,
that is calculated before data. The data operands arrive sequentially
for all the Stores, causing a high concurrency on the use of memory
for the original WaveScalar. So, in this two particular situations
there is almost no more concurrency to be extracted with the use of
Transactional WaveCache. For the \texttt{MATRIX-STORES-MIN} and
\texttt{MATRIX-STORES-MIN-DEP} the maximum speedups where 33.1\% and
24\%, showing that when the initialization loop is followed by massive
use of the initialized data, the performance gets better.

The \texttt{VECTOR-FULL-DEP} application achieved a speedup of 139.7\%
for Transactional WaveCache, that was higher that the speedup of
Decoupled Stores (136.8\%). There were also a significant amount of
hazards, specially for an infinite Speculation Window. This shows that
Transactional WaveCache improve performance even in the presence of
hazards.

\begin{table}[p]
\scriptsize
\centering
\begin{tabular}{|l|c|c|c|c|c|c|c|c|c|c|c|c|} \hline
Application & 
\multicolumn{3}{|c|}{MATRIX} & \multicolumn{3}{|c|}{MATRIX-DEP} 
& \multicolumn{3}{|c|}{MATRIX-STORES} & \multicolumn{3}{|c|}{MATRIX-STORES-DEP} \\
\cline{1-13}
Simulation & RAW & WAR & WAW  & RAW & WAR & WAW  & RAW & WAR & WAW  & RAW & WAR & WAW\\ \hline
No TWC		& 0 & 0 & 0 & 0 & 0 & 0 & 0 & 0 & 0 & 0 & 0 & 0 \\ \hline
No TWC - DEC	& 0 & 0 & 0 & 0 & 0 & 0 & 0 & 0 & 0 & 0 & 0 & 0\\ \hline
TWC 2		& 0 & 0 & 0 & 1 & 0 & 0 & 0 & 0 & 0 & 0 & 0 & 0\\ \hline
TWC 3		& 0 & 0 & 0 & 1 & 0 & 0 & 0 & 0 & 0 & 0 & 0 & 0\\ \hline
TWC 5		& 0 & 0 & 0 & 1 & 0 & 0 & 0 & 0 & 0 & 0 & 0 & 0\\ \hline
TWC 10		& 0 & 0 & 0 & 1 & 0 & 0 & 0 & 0 & 0 & 0 & 0 & 0\\ \hline
TWC 20		& 0 & 0 & 0 & 1 & 0 & 0 & 0 & 0 & 0 & 0 & 0 & 0\\ \hline
TWC 30		& 0 & 0 & 0 & 1 & 0 & 0 & 0 & 0 & 0 & 0 & 0 & 0\\ \hline
TWC - No Limits	& 0 & 0 & 0 & 1 & 0 & 0 & 0 & 0 & 0 & 0 & 0 & 0\\ \hline
\end{tabular}

\begin{tabular}{cc}
\multicolumn{2}{c}{}
\end{tabular}

\begin{tabular}{|l|c|c|c|c|c|c|c|c|c|} \hline
Application & \multicolumn{3}{|c|}{MATRIX-STORES-MIN} & \multicolumn{3}{|c|}{MATRIX-STORES-MIN-DEP} & \multicolumn{3}{|c|}{VECTOR-FULL-DEP} \\
\cline{1-10}
Simulation & RAW & WAR & WAW  & RAW & WAR & WAW  & RAW & WAR & WAW\\ \cline{1-10}\cline{1-10}
No TWC		& 0 & 0 & 0 & 0 & 0 & 0 & 0 & 0 & 0 \\ \cline{1-10}
No TWC - DEC	& 0 & 0 & 0 & 0 & 0 & 0 & 0 & 0 & 0\\ \cline{1-10}\cline{1-10}
TWC 2		& 0 & 0 & 0 & 6 & 0 & 0 & 0 & 0 & 0\\ \cline{1-10}
TWC 3		& 0 & 0 & 0 & 3 & 0 & 0 & 0 & 0 & 50\\ \cline{1-10}
TWC 5		& 0 & 0 & 0 & 5 & 0 & 0 & 0 & 0 & 101\\ \cline{1-10}
TWC 10		& 0 & 0 & 0 & 3 & 0 & 0 & 1 & 1 & 261\\ \cline{1-10}
TWC 20		& 0 & 0 & 0 & 3 & 0 & 0 & 1 & 1 & 321\\ \cline{1-10}
TWC 30		& 0 & 0 & 0 & 3 & 0 & 0 & 0 & 3 & 358\\ \cline{1-10}
TWC - No Limits	& 0 & 0 & 0 & 3 & 0 & 0 & 0 & 236 & 179\\ \cline{1-10}
\end{tabular}
\normalsize
\caption{\label{haz} Hazards detection for each application}
\end{table}

\Section{Conclusions and Future Work}\label{conc}

The advent of multi-cores has kindled interest of novel architectures,
such as WaveScalar, a DataFlow style architecture that can run
imperative programs. We present the Transactional WaveCache, a memory
disambiguation technique for WaveScalar that allows memory operations
from different waves to execute out-of-order in a speculative
way. Initial results show that very significant speedups can be
achieved through this mechanism, even in the presence of
hazards. Therefore, we believe that our results show that aggressive
speculation on memory accesses can improve performance for WaveScalar
style architectures. Indeed, our results show benefits to be
increasing even at very deep levels of speculation.

Our results confirm previous work indicating that the WaveScalar
default memory architecture could be a limiting factor in
performance. We obtain similar results to the Decouple Store
mechanism, that performs memory disambiguation within a wave. Since
the Transactional WaveCache acts in different waves, both
contributions are orthogonal, suggesting that a combination of the two
techniques should be a subject of further study.

Our work motivates a large number of future research directions.
First, our implementation relies on prior work the Kahuna simulator
and the WaveScalar binary translator. As a next step, we plan to
improve both systems so that
the Transactional WaveCache to run larger applications. This will
allow for more complete understanding of the benefits and limitations
of our technique.

We also should not notice that our design is one point in the very
large design space that has been explored by the Transactional Memory
community. In the future, and benefitting from our experience with the
Transactional WaveCache, we plan to study alternative designs, namely
towards simplifying design complexity. We also plan to study other
user and/or low-level mechanisms to throttle parallelism, and to study
whether these data-structures could also be used to achieve
synchronization.

\bibliographystyle{frontiers2008}
\bibliography{frontiers2008}

\end{document}